\documentclass{eas}
\usepackage{graphicx,natbib}
\begin{document}

\title{Gravitational Collapse of Plasmas in General Relativity} 
\runningtitle{Gravitational Collapse of Plasmas}
\author{Paul Lasky}\address{Monash University, Australia\\
		\email{paul.lasky@sci.monash.edu.au}}
\author{Anthony Lun}\sameaddress{1}

\begin{abstract}
We provide a covariant derivation of plasma physics coupled to gravitation by utilizing the $3+1$ formulation of general relativity, including a discussion of the Lorentz force law.  We then reduce the system to the spherically symmetric case and show that all regions of the spacetime can be represented in a single coordinate system, thus revoking the need for junction conditions.  We further show that the region exterior to the collapsing region is naturally described by the charged Vaidya spacetime in non-null coordinates.
\end{abstract}
\maketitle
\section{Introduction}

The standard approach when encountering gravitational collapse problems in general relativity involves utilizing junction conditions to join separate spacetimes into one single spacetime; that is, such that the union of the metrics of the individual spacetimes forms a valid solution of the Einstein field equations.  The most common of these are the Israel-Darmois junction conditions which demand the first and second fundamental forms of the neighbouring spacetimes be continuous across the boundary.  While this paradigm has had much success (see \cite{lake87} for a review) the method is fraught with difficulties (eg \cite{kirchner04}).  We present here a qualitative overview of an alternative method which enables all regions of the spacetime to be expressed in a single coordinate system, thus alleviating the need for junction conditions.  That is, such that the entire spacetime is expressed as a single solution of the Einstein field equations.  

The difficulties associated with this method are now diverted from the junction conditions.  Instead, the primary difficulty is generating the overall spacetime itself such that arbitrary matter distributions can be prescribed throughout.  The approach we have taken involves utilizing the $3+1$ formulation of the Einstein field equations, implying the problem becomes that of an initial/free-boundary value problem.  In this way we can stipulate data on some initial spacelike Cauchy hypersurface, such that some particular matter distribution exists out to a finite radius, and vanishes continuously beyond this point.  This initial hypersurface can then be evolved such that an entire four-dimensional spacetime is established.  The ``boundary'' between the two regions of the spacetime is now simply defined as the point on any subsequent spacelike hypersurface at which the matter tends to zero.  

Due to spatial constraints we provide here only a brief qualitative review of the spherically symmetric results.  These can be found in more detail in  \cite{lasky06b,lasky07b,lasky07}.  We conclude by discussing future applications of this formalism which aim to describe systems which emit gravitational radiation.

\section{Spherical Symmetry}
We divide this section on spherically symmetric gravitational collapse into three subsections according to the matter content of the respective spacetimes.

\subsection{Perfect Fluid}
A perfect fluid is one in which the stress-energy tensor has three repeated and one distinct eigenvalue given by the isotropic pressure and energy-density respectively.  Euler's equation, together with an equation of state, then implies that the lapse function (which governs the temporal evolution of successive spacelike hypersurfaces) can be deduced in terms of just the energy-density.  The entire system can then be reduced to two coupled evolution equations which govern the two remaining metric coefficients, which are the mass and energy functions (this energy function is equivalent in the appropriate limit to the energy function of the Lemaitre-Tolman-Bondi dust spacetimes).  To solve these two evolution equations, one is required to prescribe an initial energy-density and it's time rate of change.  The equation of state then implies the initial pressure is known and the definition of the mass function implies it is also known initially.  Moreover, the evolution equation for the mass can be rearranged such that the initial energy function is known.   

The way in which this initial data is prescribed is entirely dependant on the problem being studied.  We consider mainly the case of the gravitational collapse of some object which has a finite radial extent, a point we denote by $r_{\star}$.  Beyond this radius, i.e. for $r>r_{\star}$, the initial energy-density distribution falls to zero such that there is vacuum extending out to spacelike infinity.  By then solving the coupled evolution equations using this initial data, one solves the Einstein field equations for the entire spacetime.  As expected, it can be shown that for any $r\ge r_{\star}(t)$ on any subsequent constant time slice, the spacetime is diffeomorphic to the Schwarzschild spacetime.

\begin{figure}
	\begin{minipage}{1\textwidth}
		\begin{center}
		\includegraphics[height=4cm,width=4cm]{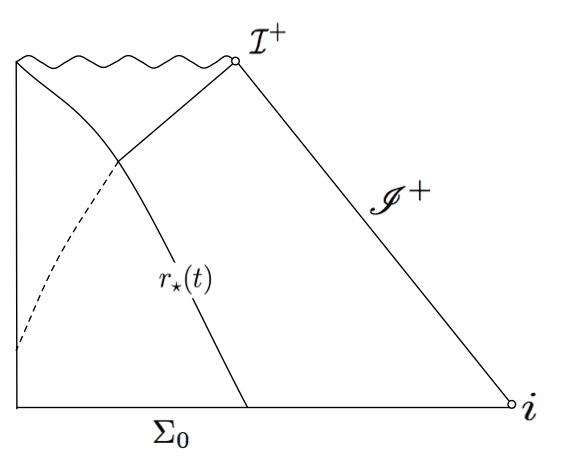}\hspace{1.5cm}
		\includegraphics[height=4cm,width=4cm]{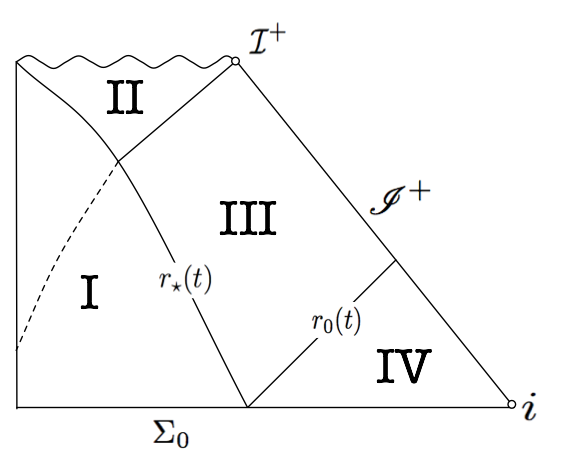}
		\caption{\label{PF} Compactification diagrams for the collapse of a perfect fluid and a radiating general fluid respectively}
	\end{center}
	\end{minipage}
\end{figure}

The left hand diagram in figure \ref{PF} shows the compactification diagram of the collapse of a perfect fluid spacetime which collapses to form a black hole.  The initial spacelike hypersurface, $\Sigma_{0}$, is where the initial data is prescribed with $r_{\star}(0)$ being the point at which the matter falls to zero.  As the system evolves, the worldline given by $r_{\star}(t)$, defined by the point on any constant time slice where the energy-density vanishes, is traced out until it reaches $r=0$.  At this point a singularity has formed, and the remaining spacetime is simply schwarzschild.  The dashed line represents the apparent horizon which emanates from $r=0$ at some finite time until it crosses $r_{\star}(t)$, at which point it becomes an event horizon.

It is shown in \cite{lasky06b} that the entire spacetime represented in the left hand diagram of figure \ref{PF} is now given by a single metric which is a solution of the Einstein field equations.

\subsection{General Fluid}
One can generalize the above to include contributions of energy flux and anisotropic stresses.  The generalization of the method is rather trivial, however the results are slightly more interesting.  We again define an initial hypersurface with some initial matter distribution that vanishes for some $r\ge r_{\star}$.  The evolution of the system is rather more cumbersome, however one can ascertain that the Vaidya spacetime is now included as a subset of the general fluid equations.  

Thus, in this model we prescribe an initial spacelike hypersurface with some matter distribution out to some finite radius $r_{\star}$.  Allowing this system to evolve under the correct conditions implies the system can radiate null particles in the form of a Vaidya region (region $III$ of right hand diagram of figure \ref{PF}).  Thus, on any subsequent constant time slice of the spacetime, one has an interior general fluid region (region $I$) reaching out to $r_{\star}$, at which point the equations degenerate into those describing a region of Vaidya spacetime.  Moving further out along this slice, one reaches the furthest point to which the radiation has been emitted, denoted $r_{0}(t)$, beyond which one recovers the Schwarzschild spacetime (region IV).  Again, an apparent horizon will form, and when this crosses $r_{\star}$ will become an event horizon, implying region $II$ is simply the Schwarzschild spacetime.  We have therefore described all regions of this spacetime in terms of a single solution of the spacetime.


\subsection{Plasmas}
Under the same formalism as described above one can express solutions to the Einstein-Maxwell field equations governing all regions of gravitational collapse.  This is a rather simple case as one is only required to perform specific mappings of the stress-energy terms to include a term governing the charge distribution.  These equations, like the general fluid equations, contain as a subset both the charged Vaidya and the Reisner-Nordstom spacetimes in the appropriate limits.  Therefore, once again one can find suitable solutions of the equations such that all regions of a collapsing spacetime are described in a single coordinate system, and are hence a single solution of the Einstein-Maxwell field equations, analogously to the two previous sections.

\section{Non-Spherical Symmetry}
This method has proven extremely fruitful in the descriptions of various fluids in the spherically symmetric regime.  The next major goal of this line of work is to establish gravitational collapse models which allow for the emission of spherical gravitational radiation.  We believe this method may be extendable to systems of equations which describe matter filled, non-spherically symmetric spacetimes, which reduce in the vacuum limit to known spacetimes.  For instance, an interior dust solution which contains the Robinson-Trautman (RT) spacetimes as a subset.

While this task is obviously non-trivial, the future looks promising as in some ways the RT spacetimes are similar in structure to the Vaidya spacetime.  Therefore, in theory it is a case of following a similar routine achieved for the Vaidya spacetime with the added degrees of freedom associated with the shapes of the two-spaces.  One could then establish an initial spacelike hypersurface which contained this interior RT dust solution which degenerates into a the Schwarzschild spacetime for all $r\ge r_{\star}$.  That is the shape of the two-spaces tends towards spheres at $r_{\star}$ on the initial hypersurface.  As the system evolves, the bumps and hills on the two-spheres in the interior region will radiate away, such that a region of vacuum RT opens between the dust RT and the Schwarzschild region, akin to region $III$ in the right hand diagram of figure \ref{PF}.

\bibliographystyle{astron}
\bibliography{ERE07Proc}
\end{document}